\documentclass[11pt]{article}

\usepackage{graphicx} 

\usepackage{lineno}

\title{4D reconstruction of alumina laser melt pools at 25 kHz via operando
X-ray multi-projection imaging}

\author{
Lars Witte$^{a}$,
Eliot Jermann$^{b, c}$,
Zhe Hu$^{d}$,\\
Zisheng Yao$^{d}$,
Eleni Myrto Asimakopoulou$^{e}$,
Julia Katharina Rogalinski$^{d}$,
Yuhe Zhang$^{d, f}$,\\
Kim Nygård$^{e}$,
Malgorzata G. Makowska$^{b, g}$,
Markus Bambach$^{a}$,\\
Mohamadreza Afrasiabi$^{a,*}$,
Pablo Villanueva-Perez$^{d}$\\[1ex]
\textit{$^{a}$Advanced Manufacturing Lab, ETH Z\"urich, Z\"urich 8092, Switzerland}\\
\textit{$^{b}$Laboratory for Nuclear Materials, Paul Scherrer Institut, Villigen, Switzerland}\\
\textit{$^{c}$LUXS Laboratory for Ultrafast X-ray Sciences, Institute of Chemical Sciences and Engineering, }\\
\textit{\'Ecole Polytechnique F\'ed\'erale de Lausanne (EPFL), CH-1015 Lausanne, Switzerland}\\
\textit{$^{d}$Synchrotron Radiation Research and NanoLund, Lund University, Lund, Sweden}\\
\textit{$^{e}$MAX IV Laboratory, Lund University, Lund, Sweden}\\
\textit{$^{f}$Medical Radiation Physics, Lund University, Lund, Sweden}\\
\textit{$^{g}$Laboratory for Synchrotron Radiation and Femtochemistry, }\\
\textit{Paul Scherrer Institut, 5232 Villigen PSI, Switzerland}\\[1ex]
\textit{*~Corresponding author:  afrasiabi@ethz.ch (M. Afrasiabi)}
}

\date{}

\begin{document}

\maketitle

\begin{abstract}
Advancing additive manufacturing, e.g., laser powder-bed fusion (LPBF), requires resolving rapid processes such as melt-pool dynamics and keyhole evolution in 4D (3D + time). Operando X-ray tomography is a state-of-the-art approach for 4D characterization, but its temporal resolution is fundamentally constrained by the sample rotation speed, limiting achievable 4D imaging rates and preventing the resolution of these fast phenomena. Here we present rotation-enabled X-ray Multi-Projection Imaging (rotation-XMPI), which captures three angularly resolved projections per time step and thereby decouples temporal resolution from the sample rotation speed. Combined with a self-supervised deep-learning reconstruction framework for multi-angle inputs, rotation-XMPI enables high-fidelity 4D imaging at unprecedented speed. We demonstrate the approach in an operando alumina laser-remelting experiment at MAX IV using three beamlets combined with 25 Hz sample rotation. Rotation-XMPI resolves melt-pool morphology and keyhole evolution; in contrast, conventional and limited-angle tomography remain rotation-limited, and motion blur prevents resolving these dynamics. Overall, rotation-XMPI delivers a 250-fold increase relative to state-of-the-art melt-pool imaging, effectively achieving 25,000 reconstructed volumes per second. This method establishes a practical route to scalable ultrafast 4D imaging for additive manufacturing and other materials processes.
\end{abstract}

\textbf{Note:} Supplementary materials, movies, datasets, and code will be made available upon publication of the peer-reviewed article.\\[1ex]


\label{sec:intro}
\par Additive manufacturing (AM) enables the production of intricate, high-complexity parts that are difficult or impossible to create using conventional machining. Many AM techniques are governed by highly localized, transient phenomena that evolve over short time and length scales, making them challenging to observe directly during fabrication. Laser powder bed fusion (LPBF) is among the most widely used AM processes, in which a scanning laser beam selectively fuses powder to the part, building it layer by layer. The process is governed by rapid three-dimensional melt-pool dynamics that control defect formation and process stability \cite{khairallah2016laser, khairallah2020controlling, gan2021universal, afrasiabi2023modelling, ullah2025additive}.

Laser melt-pool simulations provide insight into melt-pool dynamics and can support process optimization \cite{muther2025identifying, luthi2023adaptive}, but they are subject to modeling errors and computational limitations \cite{moges2019review}. Experimental approaches are therefore needed to validate such models. Among existing experimental techniques, X-rays offer the penetration needed to observe these processes under operando and in situ conditions without disturbing them. While two-dimensional X-ray imaging has revealed rapid melt-pool evolution, keyhole oscillations, and pore dynamics at frame rates above 10 kHz \cite{parab2018ultrafast, martin2019dynamics, martin2022laser}, direct access to the evolving internal 3D melt-pool morphology (provided, for example, by operando X-ray tomography) is required to understand these processes and validate numerical simulations.

Conventional operando tomography is a state-of-the-art 4D (3D+time) X-ray imaging technique~\cite{garcia2021tomoscopy, garcia2023x}. However, it is constrained by rotation-dependent acquisition: each tomogram requires a substantial angular range of projection directions, typically 180° for parallel-beam projections. This couples temporal resolution to the rotation speed and limits reconstruction rates for fast processes. Increasing rotation speed is often undesirable in operando studies because it can introduce mechanical perturbations, and centrifugal forces may alter the dynamics. As a result, despite synchrotron flux enabling short exposures and high frame rates, full 4D reconstructions for LPBF have been limited to $\sim$100 volumes per second (vps) \cite{muther2025identifying, makowska2023operando}. Even state-of-the-art operando tomography at $\sim$1000 vps~\cite{garcia2021tomoscopy} remains insufficient to resolve key transient melt-pool dynamics in many commercially relevant systems. Alternative 4D X-ray approaches are thus needed to study laser melting and to gain insight into melt-pool dynamics, morphology, and defect-formation mechanisms.

X-ray multi-projection imaging (XMPI) offers an alternative route to fast 4D imaging by recording multiple projection angles simultaneously using a multibeam X-ray configuration~\cite{hoshino2011stereo,mokso2015StereoDual,Villanueva-Perez2018XMPI,Duarte2019CDIstereo,voegeli2020multibeamWhite}. Multiple beamlets are created using crystals and captured by separate detectors, yielding sparse but instantaneous multi-view data and partially decoupling temporal resolution from sample rotation. Previous XMPI demonstrations have achieved effective frame rates from beyond kHz up to the mHz range \cite{asimakopoulou2024kHzXMPI, voegeli2024multibeamSiArray, villanueva2023megahertz}. However, to access dynamics at 10 kHz and beyond, as required for melt-pool studies, only a very sparse set of projections (often three or four) is currently available ~\cite{Bellucci2024CrystalsXMPI}, which poses a reconstruction challenge.

Reconstruction of 4D attenuation or phase-shift fields from sparse and/or limited-angle projections is a well-known problem in the imaging community. Standard 3D reconstruction approaches, such as filtered back-projection, perform poorly under these conditions, motivating iterative and learning-based algorithms that capture spatiotemporal priors. Direct sparse/limited-angle 4D reconstruction in a voxel representation can be computationally demanding and memory-intensive \cite{goethals2022dynamic, mohan2015timbir, kazantsev20154d}. Recent approaches include event-based dynamic reconstruction (DYRECT) \cite{goethals2025dyrect} and neural radiance-field-based representations \cite{mildenhall2021nerf, zhang20254d, reed2021dynamic}. Among these, X-Hexplane~\cite{hu2025super} achieves efficient, high-quality 4D reconstruction by encoding spatial and temporal features in a shared low-dimensional Hexplane representation~\cite{cao2023hexplane}. Nevertheless, such an approach has not yet been used in conjunction with XMPI.

In this paper, we demonstrate rotation-XMPI as a method for high-speed sparse 4D reconstruction of laser melt pools. The method combines XMPI with three beamlets, continuous but slow sample rotation, and a 4D self-supervised deep-learning reconstruction to achieve 4D reconstructions with temporal resolution primarily limited by the detector frame rate. The presented operando investigations of alumina melt-pool dynamics at MAX IV demonstrate the capabilities of rotation-XMPI and its potential for studies of other processes that are not accessible with conventional microscopy. When applied to laser-based AM, rotation-XMPI captures melt-pool morphology and its evolution at 25,000 vps, corresponding to a 250× increase in effective reconstruction rate compared with prior LPBF operando tomography using state-of-the-art 4D X-ray imaging methods~\cite{muther2025identifying, makowska2023operando}. More broadly, combining rotation-XMPI with self-supervised 4D reconstruction provides a clear pathway to high-speed 4D imaging of transient phenomena across a wide range of materials and engineering systems.

\section{Results}
\label{sec:results}

\subsection{Reconstruction of laser-induced melt pools with rotation-XMPI}
\par
\begin{figure}
    \centering
    \includegraphics[width=1\linewidth]{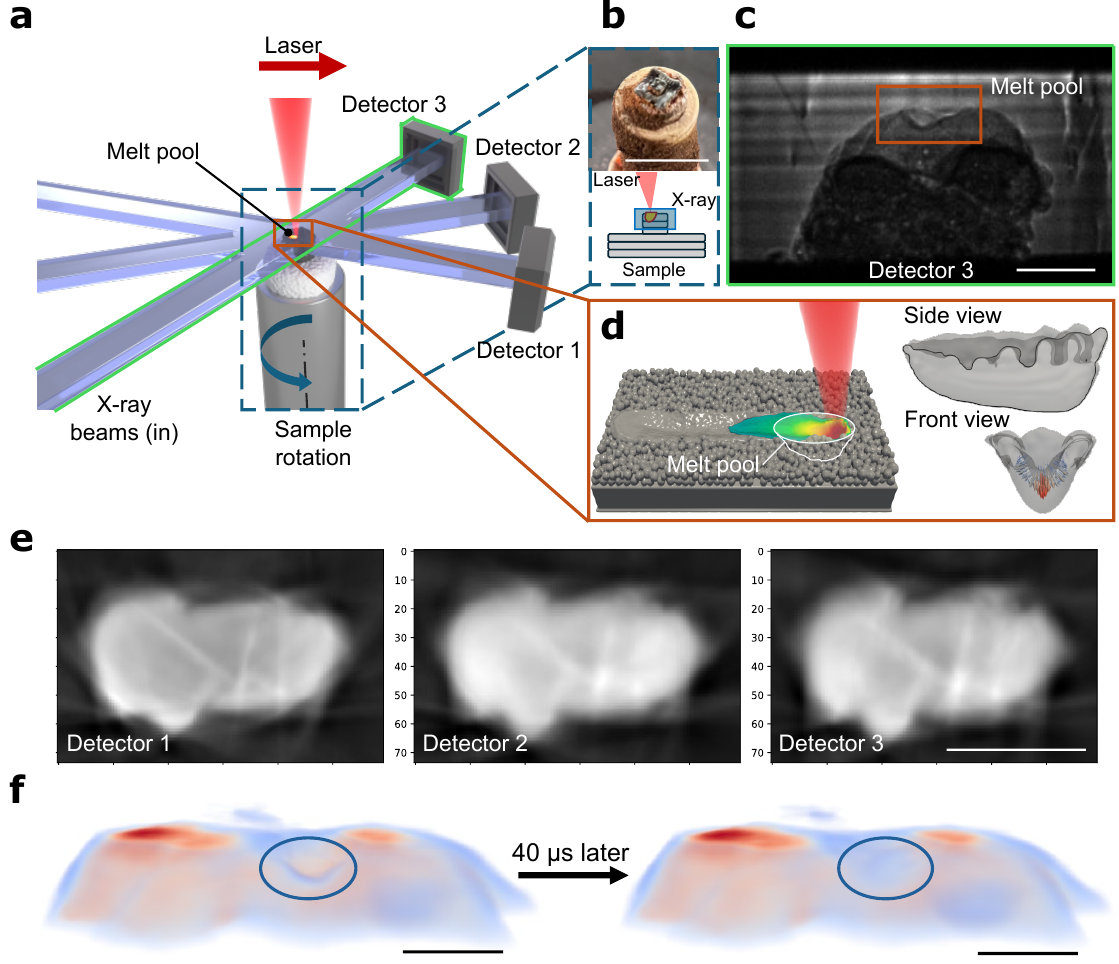}
    \caption{\textbf{Acquisition setup and reconstruction results of laser remelting.} \textbf{a} Schematic of the XMPI data-acquisition setup. Three X-ray beamlets are shown in blue and the laser in red. The sample is mounted on a rotating holder. \textbf{b} Photograph of the sample mounted on the holder (background grayed out) and a schematic of the sample during remelting. \textbf{c} Raw projection without flat-field correction, with the melt pool visible. \textbf{d} Simulated melt-pool behavior and morphology, adapted from Refs. \cite{luthi2025multi,afrasiabi2022effect}; unlike the powder-bed simulation, the experiment remelts solid material without a powder layer. \textbf{e} Horizontal orthoslices reconstructed from single-detector data for the static initial sample state (each detector reconstructed independently).   \textbf{f} Two time steps from the rotation-XMPI reconstruction, separated by 40~\textmu{}s. Scale bars: 6~mm in \textbf{b}; 200~\textmu{}m in \textbf{c}; 200~\textmu{}m in \textbf{e}; 100~\textmu{}m in \textbf{f}.}
    \label{fig:setup}
\end{figure}

 Rotation-XMPI, combined with an adapted X-Hexplane algorithm, was employed to reconstruct the 4D evolution of an alumina melt pool during laser remelting. The rotation-XMPI setup (Fig.~\ref{fig:setup}a) utilized three synchronized X-ray projections of the sample (Fig.~\ref{fig:setup}b), each recorded with a detector operating at a frame rate of 25 kHz (Fig.~\ref{fig:setup}c). The sample was mounted on a rotation stage running at 25 rotations per second, corresponding to an effective rate of 50~vps in conventional operando tomography.

At every imaging time step, three projections were acquired simultaneously along the beam directions indicated in Fig.~\ref{fig:setup}a, with angular separations of $0^\circ$, $30.7^\circ$, and $47.7^\circ$ relative to the beam forming the projection on detector~1. To investigate the evolution of the alumina melt pool (schematically illustrated in Fig.~\ref{fig:setup}d), a 75~W laser beam was temporally modulated with a period of $T = 500\,\mbox{\textmu{}s}$ and an on-time of $\tau_{\mathrm{on}} = 150\,\mbox{\textmu{}s}$. The beam was focused using an F-theta lens and scanned at 50~mm/s across the sample in the sample's coordinate system reference frame (Fig.~\ref{fig:setup}b), generating a straight line laser track. The melt pool is distinguishable in the projections of all three detectors (Supplementary Fig.~S1).

Before volume reconstruction, the projection data underwent a multi-step preprocessing procedure. The three projections (an example from detector~3 is shown in Fig.~\ref{fig:setup}c) were geometrically aligned, intensity-homogenized across detectors, and processed to suppress imaging artifacts. A detailed description of the preprocessing workflow is provided in Methods: Projection data preprocessing and Supplementary Movie~1. After preprocessing, inter-detector consistency was assessed by reconstructing conventional static tomograms independently from each detector using 500 fully processed projections of the initial sample state spanning a $180^\circ$ rotation (Fig.~\ref{fig:setup}e). These reconstructions exhibited good agreement, confirming the effectiveness of the preprocessing pipeline.

Subsequently, 4D reconstructions of the laser remelting process were obtained using the adapted X-Hexplane algorithm. Two consecutive reconstruction time steps are shown as volumetric renderings in Fig.~\ref{fig:setup}f, where the development of the laser-induced indentation is clearly identifiable within the laser interaction zone. The reconstructed 4D domain comprised $200 \times 200 \times 16$ voxels over 700 time steps, with an isotropic voxel size of $4 \times 4 \times 4\,\mbox{\textmu{}m}^3$ and an effective temporal spacing of 40~\textmu{}s between successive volumes. This corresponds to a field of view of $0.8 \times 0.8 \times 0.064~\mathrm{mm}^3$ over a total duration of 28~ms. The complete 4D reconstruction was obtained in under one hour on a single NVIDIA GeForce RTX~4070~Ti GPU using 10.5~GB of VRAM (loss evolution in Supplementary Fig.~S4). A representative horizontal slice through the reconstructed volume sequence is shown in Supplementary Movie~2.

\subsection{Comparison of rotation-XMPI with state-of-the-art 4D reconstruction techniques}
To quantify the benefit of rotation-XMPI for 4D X-ray imaging, the reconstructed 4D datasets were compared against (i) conventional operando tomography and (ii) Super Time-Resolved Tomography (STRT)~\cite{hu2025super}, a state-of-the-art approach for enhancing temporal resolution beyond classical tomographic methods.

\begin{figure}
    \centering
    \includegraphics[width=1\linewidth, trim=0 9.1cm 0 0, clip]{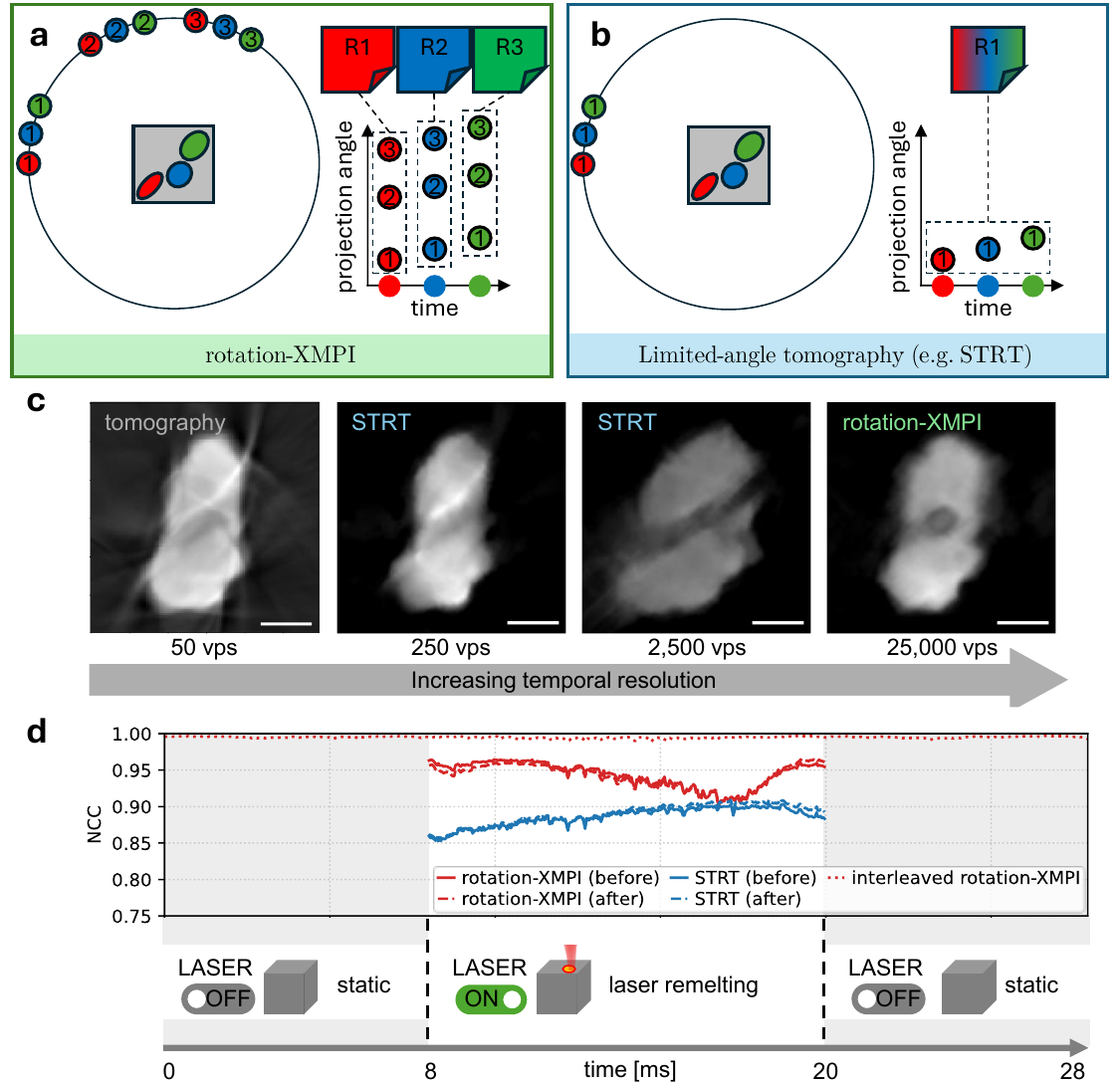}
    \caption{\textbf{Comparison of imaging and reconstruction approaches.}
    \textbf{a,b} Schematics of rotation-XMPI and limited-angle tomography in the sample's frame of reference. Circled numbers denote the detector positions, and colors encode acquisition times (red/blue/green); the center depicts schematic melt-pool positions at three times. R1–R3 indicate the corresponding reconstructions.
    \textbf{c} Comparison of reconstructions using conventional tomography (Gridrec), limited-angle (STRT), and rotation-XMPI algorithms. Reconstructions are shown as horizontal slices and are cropped to the sample for clarity. Scale bars: 100~\textmu{}m.
    \textbf{d} Normalized cross-correlation (NCC) between dynamic reconstructions (rotation-XMPI and STRT) and static reference volumes acquired before and after laser remelting. NCC between two independent rotation-XMPI reconstructions (even/odd frame subsets) is also shown as a robustness evaluation.}
    \label{fig:STRTvsXMPI}
\end{figure}

In rotation-XMPI, three synchronized projections (from detectors 1--3) were acquired at each imaging time step and jointly reconstructed with the adapted X-Hexplane algorithm (Fig.~\ref{fig:STRTvsXMPI}a). Both conventional operando tomography and STRT relied on a single projection per time step (Fig.~\ref{fig:STRTvsXMPI}b). Detector~1 was selected for these single-detector reconstructions because it provided the sharpest and least obstructed view of the melt pool. Detectors~2 and~3 were excluded due to blurring from Laue diffraction and the Borrmann triangle effect~\cite{rogalinski2025time, authier2006dynamical} and due to transient occlusion by a powder particle during remelting, respectively.

For conventional operando tomography, each volume was reconstructed with the Gridrec algorithm~\cite{dowd1999developments, gursoy2014tomopy} with all 500 projections spanning $180^\circ$. When the process evolves faster than the time required to acquire a tomogram, temporal blurring occurs. If the sample rotation speed cannot be increased, limited-angle reconstruction approaches, such as STRT, can be employed. This involves reconstructing each time step from a limited angular range, thereby trading angular coverage for temporal sampling. Here, temporal enhancement factors ranging from 5 to 500 were evaluated compared to conventional tomography. For example, an enhancement factor of 5 corresponds to $36^\circ$ (100 projections) per reconstruction volume, whereas an enhancement factor of 500 corresponds to a single projection per reconstruction volume. Increasing the enhancement factor, therefore, improves temporal sampling but reduces angular coverage and projection count, which can degrade spatial fidelity through angular undersampling.

Reconstructions obtained during laser processing are shown in Fig.~\ref{fig:STRTvsXMPI}c, ordered by reconstructed volumes per second. The conventional tomographic reconstruction (500 projections over $180^\circ$) exhibits severe temporal blurring: over the 500-frame acquisition window, the laser modulation cycle elapses 40 times, smearing the melt pool over multiple modulation periods and preventing temporal resolution of the cycle. STRT exhibits the expected trade-off. At high temporal enhancement factors, angular undersampling leads to pronounced reconstruction artifacts, whereas at lower enhancement factors, the improved angular coverage comes at the expense of temporal smearing. Across all tested enhancement factors, the melt pool appears broadened and low in contrast, and a well-defined keyhole cannot be clearly resolved.

This limitation is consistent with the ratio between the process time scale and the rotation rate. The laser modulation period is $T = 500\,\mbox{\textmu{}s}$, corresponding to 12.5 imaging frames at the effective temporal spacing of 40~\textmu{}s. STRT reconstructions that aggregate 10 imaging frames per reconstruction volume (2,500 vps), therefore, insufficiently sample the melt-pool cycle while also operating under a severely limited angular range, which compromises the spatiotemporal fidelity. In contrast, rotation-XMPI reconstructs one volume per imaging time step by combining the three simultaneous projection views with the maximum angular difference up to $47.7^\circ$, effectively matching the reconstruction frame rate to the acquisition frame rate and resolving melt-pool and keyhole evolution.

To quantitatively assess the reconstruction fidelity of rotation-XMPI, normalized cross-correlation (NCC) was computed (Fig.~\ref{fig:STRTvsXMPI}d; Supplementary Fig.~S5). Because no dynamic 4D ground truth is available and the evolving melt region occupies only a small fraction of the total sample volume, static tomographic reconstructions acquired before and after remelting were used as surrogate reference volumes. For each static state, one 3D reconstruction per detector was computed and the three detector-specific volumes were averaged to form a single reference volume. Rotation-XMPI exhibits consistently higher agreement with these references than STRT when STRT is configured with one imaging frame per reconstruction volume (temporal enhancement factor of 500). Reconstruction robustness was further evaluated by independently reconstructing two temporally interleaved projection subsets (odd and even frames) and computing the NCC between the resulting reconstructions, which showed close agreement.

\subsection{Melt pool study}
\begin{figure}
    \centering
    \includegraphics[width=1\linewidth, trim=0 7.8cm 0 0, clip]{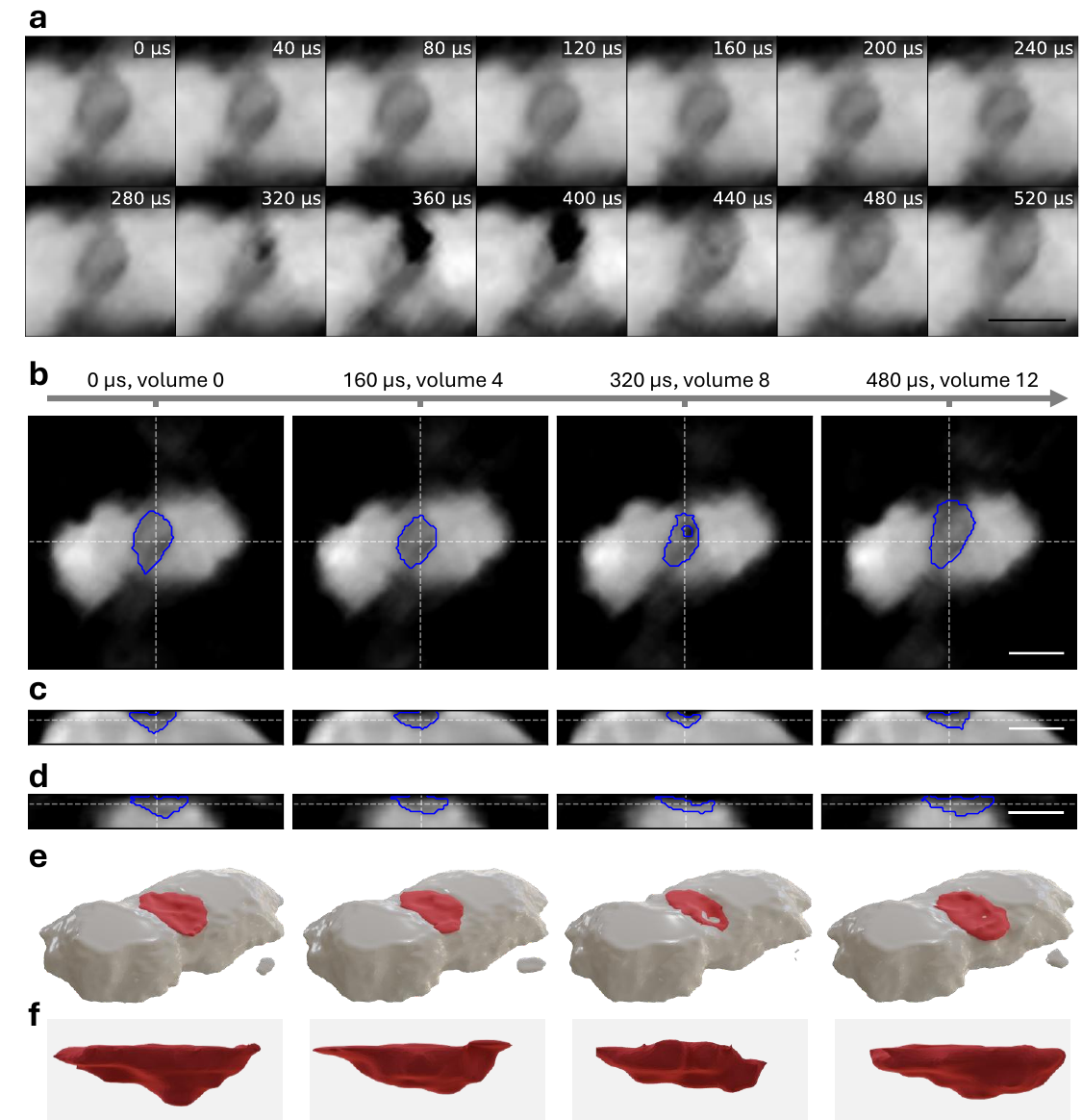}
    \caption{\textbf{Melt pool segmentation.}
    \textbf{a} Sequence of 14 cropped horizontal slices illustrating the evolution of the keyhole and melt pool; the laser scans from bottom-left to top-right in each panel. \textbf{b} Horizontal slices with the segmented melt pool overlaid; the four time points are separated by 160~\textmu{}s. \textbf{c, d} Vertical slices with the segmented melt pool overlaid. \textbf{e} 3D rendering of the segmented melt pool together with the solid sample. \textbf{f} Side-view rendering of the segmented melt pool. Scale bars: 100~\textmu{}m in \textbf{a--d}.}
    \label{fig:segmentation}
\end{figure}

Figure~\ref{fig:segmentation}a visualizes the melt-pool evolution over a time interval slightly exceeding one full laser modulation period. In the reconstructed phase-shift field, solid alumina exhibits the highest phase shift, air-filled regions show negligible phase shift, and molten material within the melt pool displays intermediate phase shift. The sequence spans 520~\textmu{}s and begins and ends with the laser turned off. At the beginning of the sequence, the previously molten region contracts as solidification proceeds, during which the melt pool remains stationary. Upon laser activation, a keyhole forms and deepens, appearing as a low-phase-shift indentation within the melt pool. The melt-pool position shifts between consecutive frames, consistent with the imposed laser scanning speed. After the laser is turned off, the keyhole collapses and the molten material flows back. These dynamics are resolved with a temporal sampling interval of 40~\textmu{}s, enabling continuous 3D tracking of melt-pool morphology and keyhole evolution.

Based on the reconstructed volumes, manual melt-pool segmentation was performed, with results shown in Fig.~\ref{fig:segmentation}b--f. The melt pool is widest and deepest in the first frame and progressively decreases in size in subsequent frames. A laser-induced indentation driven by vapor pressure is observed in frame~8. In the final frame, the melt pool again exhibits an increased size. Owing to residual reconstruction blur, the segmentation is inherently imprecise and relies on systematic inspection across multiple time steps, slice depths, and slice orientations. The distinction between liquid and solid phases is more reliable than the segmentation of the liquid--air interface, as the latter is affected by blurring at the adjacent solid--air boundary. The solid material was segmented automatically using thresholding of the phase-shift field. A rendering of the segmented melt pool over three laser modulation periods is shown in Supplementary Movie~3.

\section{Discussion}
\label{sec:discussion}
The present study demonstrates that rotation-XMPI, combined with an adapted X-Hexplane reconstruction, enables high-speed 4D imaging of laser-induced melt pools. By synchronizing three projection angles with continuous sample rotation and performing joint 4D reconstruction, a 250-fold increase in effective temporal resolution relative to prior operando LPBF tomography is achieved. This capability allows melt-pool morphology, keyhole formation, and collapse to be resolved at time scales that were previously accessible only with two-dimensional imaging techniques.

The comparison with conventional tomography and STRT highlights the advantages of the proposed framework for investigating such fast, transient phenomena. Conventional tomography requires a large angular range per reconstructed volume and therefore exhibits pronounced temporal smearing when applied to dynamics that evolve much faster than the achievable rotation speed, such as melt-pool evolution. STRT partially alleviates this limitation by exploiting an efficient 4D representation; however, in the present setting it remains constrained by the extremely sparse angular coverage available per time step. For temporal enhancement factors sufficiently large to resolve the laser modulation period, the angular range becomes too limited to reconstruct the static sample and melt pool with adequate spatial fidelity. In contrast, rotation-XMPI acquires multiple projection angles simultaneously and employs an adapted X-Hexplane representation to couple all time steps within a single 4D optimization. While the data term remains local in time, the shared representation and spatiotemporal regularization propagate information across the sequence, enabling frame-rate-matched 4D reconstructions from highly sparse projection data.

The comparison between STRT and rotation-XMPI isolates the benefit of additional projection directions per time step from differences in acquisition hardware. Both methods are evaluated using XMPI-acquired projections, such that the comparison reflects the value of multi-view information rather than a full end-to-end comparison between an optimized single-beam system and the rotation-XMPI setup. Nonetheless, this comparison is instructive: STRT has been reported to provide a 10--20$\times$ temporal enhancement relative to $180^\circ$ operando tomography, whereas in the present experiment rotation-XMPI achieves a 500$\times$ enhancement relative to the 25~Hz sample rotation speed, corresponding to a 25--50$\times$ larger improvement than previously reported for STRT.

The reconstructions also reveal practical constraints of the current implementation. The accessible field of view and spatial resolution are limited by the multibeam geometry, detector performance, and the chosen voxel grid, and the reconstruction quality is sensitive to imaging artifacts introduced by the beam splitting setup. A dedicated preprocessing pipeline was therefore essential (see Methods section). Geometric alignment, jitter correction, denoising, and Paganin filtering were used to harmonize the three beamlets and to suppress artifacts before reconstruction. While these steps substantially improved reconstruction quality, they also introduced additional complexity and potential sources of bias, suggesting that further optimization and partial automation of the preprocessing workflow could yield additional gains in robustness and reproducibility.

The present validation relies on normalized cross-correlation with static reference volumes and internal consistency checks. A more comprehensive assessment using realistic synthetic XMPI datasets, combined with uncertainty quantification, would provide deeper insight into reconstruction accuracy and potential failure modes. In this context, incorporating physics-based priors and differentiable models of diffraction into the reconstruction framework represents a promising direction for further improving reconstruction fidelity~\cite{yao2025physics-informed4D}.

In conclusion, rotation-XMPI is introduced and demonstrated as a practical route to ultrafast 4D imaging in otherwise rotation-limited operando experiments. By combining three projection directions per imaging time step with continuous sample rotation and jointly reconstructing the 4D phase-shift field using an adapted, self-supervised X-Hexplane framework, the effective temporal sampling shifts from being primarily rotation-limited to detector-frame-rate-limited, yielding 25,000 reconstructed volumes per second with a temporal spacing of 40~\textmu{}s. 

This represents a 250-fold increase in reconstructed-volume rate relative to prior operando LPBF tomography. In alumina laser remelting experiments at MAX~IV, this advance enables direct 3D tracking of melt-pool morphology and keyhole formation and collapse with substantially reduced motion blur, whereas conventional tomography and limited-angle STRT baselines using a single detector fail to resolve the melt pool. Further improvements in spatial and temporal resolution are expected from advances in beamline stability, multibeam optics, detector performance, and reconstruction algorithms, particularly for delineating challenging interfaces such as the liquid--air boundary. From an application perspective, the approach can be extended from alumina remelting to LPBF of ceramics, LPBF of metallic alloys, and other technologically relevant materials systems. Moreover, the proposed approach is particularly interesting for direct investigation of the interaction of laser with matter, which requires resolving the melt pool shape in 4D,  enabling exploring various laser modulations, for instance in time (pulsed lasers), as well as geometrical beam-shaping strategies.

\section{Methods}
\label{sec:methods}
\subsection{LPBF of alumina}
The alumina sample was produced with a custom-built 3D printer consisting of a control system (Aerotech controllers and computer), a cooling system, a 500 W infra-red fiber laser (IPG Photonics), a laser galvo-scanner (Aerotech GL4), an F-theta lens, and a rotating build plate. The detailed description of the original setup can be found in \cite{makowska2023operando}. The setup used in this work is based on that design but includes small updates such as a new laser and a controlling computer. Its unique feature is its ability to print complex geometries during build plate rotation through precise synchronization of the rotation and the galvo scanner. The laser is a continuous-wave system that can be modulated by an external trigger to define pulses over a defined period. This functionality was used to achieve the average laser power output required to print alumina.

The samples were printed layer by layer on an alumina foam substrate with manual recoating prior to the imaging experiment. The composition of the sample precursor powder is aluminum oxide (Al$_2$O$_3$) doped with $5.2~vol\%$ iron oxide in the form of magnetite. As shown in Fig.~\ref{fig:setup}b, the sample is composed of two parts: a larger base, which provides stability during printing, and a rectangular column built on top of it with approximate dimensions 2 x 0.5 x 0.5 $~\mathrm{mm}^3$. The cross-sectional dimensions of 0.5 x 0.5 $~\mathrm{mm}^2$ were defined according to the imaging FOV. After printing, the powder was removed from the sample to perform remelting experiments consisting of scanning a laser track through the sample while imaging with the XMPI setup. The process parameters used for the remelting experiment were as follows: laser scanning speed of 50 mm/s, a modulated laser power with pulses of 150~\textmu{}s over a period of 500~\textmu{}s targeting a theoretical average power of 22.5 W, and a laser spot size of about 140~\textmu{}m in diameter.

\subsection{Projection data acquisition}
The data-acquisition overview is shown in Fig.~\ref{fig:setup}. Data were acquired with an XMPI setup \cite{rogalinski2025time} on the ForMAX beamline~\cite{nygaard2024formax} at the MAX IV synchrotron using 16.6~keV photon energy. The beam was split into three beamlets using silicon and germanium crystals, which then passed through the sample. The angles between the beamlets were $30.7^\circ$ (detectors 1-2) and $17^\circ$ (detectors 2-3). After passing through the sample, the projections were captured by three detectors at a 25~kHz frame rate and with 4~\textmu{}m pixel size.  Each detector was equipped with an optical imaging system consisting of a Photron Nova~S16 camera, a 5$\times$ objective, and a 250~\textmu{}m-thick GaGG$+$ scintillator.

The alumina sample, approximately 500~\textmu{}m in diameter, was rotated at 25~Hz. Flat-field images were acquired before the experiment to enable flat-field correction. The X-ray–illuminated region in the images was approximately 480~\textmu{}m high and 840~\textmu{}m wide (120~px $\times$~210~px).

\subsection{Projection data preprocessing}
\label{sec:data_preprocessing}
Correction of imaging artifacts and alignment of the three detector views were essential for achieving high reconstruction quality. Because the detectors record beamlets with different X-ray paths and crystals, their flat-field-corrected projections contained detector-specific artifacts, whereas accurate reconstruction required the images to be as similar as possible across detectors and over time. Projection streams from the three beamlets were preprocessed to (i) remove detector- and beamlet-specific artifacts, (ii) stabilize residual temporal fluctuations, and (iii) harmonize contrast and geometry across detectors prior to joint reconstruction (Supplementary Fig.~S3). Unless stated otherwise, the same sequence of operations was applied to the full projection time series of each detector.

\textbf{Preprocessing quality reference.} Flat-field projections without the sample present were acquired. The sample was then brought into view and recorded during rotation before the laser was turned on. This enabled the acquisition of a static-sample dataset and comparison of detectors by matching projections at the same relative sample position but at different times. Projections, sinograms, and tomographic~\cite{dowd1999developments} reconstructions from the three detectors were then compared at different processing stages along the preprocessing pipeline. In these flat-field-corrected but otherwise raw projections, the detectors exhibited distinct artifact patterns. 
Specifically, detector~1 showed substantial noise and pronounced time-varying side-to-side brightness fluctuations. Detector~2 was blurrier and exhibited section-specific lateral jitter that dynamically stretched parts of the image. Detector~3 had lower noise and minimal jitter or brightness variation, but a powder particle traversed the field of view directly in the projection direction with the melt-pool region during laser processing. All detectors exhibited edge enhancement due to the propagation distance between the sample and detectors (Fresnel diffraction).

\textbf{Projection alignment.} First, approximate manual alignment of the detectors was performed based on the setup geometry. For detector~2, exhibiting section-specific lateral jitter and warping due to the Laue-splitting mechanism~\cite{rogalinski2025time}, a multi-point, piecewise displacement model was estimated using multiple vertical control sections within each frame, and a continuous warp field was obtained by interpolating the measured shifts. The algorithm is described in Algorithm~1. The multi-point jitter-correction algorithm was not applied to detectors~1 and~3 to avoid introducing additional interpolation artifacts where no pronounced jitter was observed.

\textbf{Brightness fluctuation correction.} Temporal side-to-side brightness changes in the projections cannot be captured by static flat-field correction. Instead, peripheral regions that are consistently outside the imaged sample are used to apply a simple first-order horizontal correction. For each projection at time $t$, mean intensities were measured in two fixed peripheral regions near the left and right borders, yielding time series $L_t$ and $R_t$. A global reference level $I_{\mathrm{ref}}$ was defined as the temporal average of ${L_t,R_t}$. Per-frame offsets were $\Delta L_t = L_t - I_{\mathrm{ref}}$ and $\Delta R_t = R_t - I_{\mathrm{ref}}$. A linear horizontal correction field was formed as
\begin{equation}
C_t(x,y) = \Delta L_t + \bigl(\Delta R_t - \Delta L_t\bigr) m(x),
\end{equation}
where $m(x)\in[0,1]$ increases linearly from the left to the right image boundary. The corrected projection was
\begin{equation}
I^{\mathrm{corr}}_t(x,y) = I_t(x,y) - C_t(x,y).
\end{equation}
which suppresses side-to-side fluctuations without imposing higher-order spatial changes. This correction was applied to all three detectors and reduced temporal side-to-side brightness fluctuations, which were particularly pronounced in detector~1 (Supplementary Fig.~S2a).

\textbf{Denoising.} 
To correct the detector noise at 25~kHz frame rate, we estimated for each detector independently the noise standard deviation $\sigma$ using \texttt{skimage.restoration.\allowbreak estimate\_sigma} \cite{donoho1994ideal} on a central region of interest (the middle 150 detector columns) to avoid the elevated noise near the beamlet boundaries. 
Fast non-local means denoising is then applied to the 3D projection stack of each detector (\texttt{skimage.restoration.\allowbreak denoise\_nl\_means})~\cite{darbon2008fast}. The filtering parameters were $h = 0.8\sigma$, patch size $=5$, and patch distance $=6$. Residual noise at the lateral margins outside the sample support was further reduced by applying Gaussian blurring in detector-specific background regions.

\textbf{Paganin (edge-enhancement) filtering.} Fresnel diffraction at the sample–scintillator distance introduces phase-contrast edge enhancement that degrades quantitative consistency across detectors and impairs reconstruction. To suppress these effects, Paganin phase retrieval \cite{paganin2002simultaneous} is applied to obtain phase-retrieved projections with reduced diffraction artifacts, with a modest loss of high-frequency detail accepted as a trade-off. The strength of Paganin filtering is controlled by the parameter $r = \delta/\beta$, which is nominally set by the sample material through the ratio of refractive index decrement $\delta$ to absorption index $\beta$, but was treated here as a tunable parameter and manually adjusted to optimize quality and inter-detector consistency. The Paganin parameter for detector~1 was fixed to $r = 2500$, and candidate values for detectors~2 and~3 were expressed as multiplicative factors of this reference value. For each combination of factors on detectors~2 and~3, the pipeline (Paganin filtering, background subtraction, and row equalization) was executed and the $L_2$-loss between detectors was evaluated over projections before and after laser processing ($timesteps = 0\mbox{--}1000$ and $timesteps = 2200\mbox{--}2850$), after aligning projections at matched sample angles. The factor combination [1, 1.25, 2.5] for detectors [1, 2, 3], corresponding to $r$-values [2500, 3125, 6250], minimized this discrepancy and was therefore used for all experiments (Supplementary Table~S1).

\textbf{Background removal.} After applying the final vertical detector offsets and Paganin filtering, the background is removed using an adapted Canny edge detection approach. Each column of each projection is analyzed independently: a pixel is marked as an edge candidate if its value exceeds a fixed threshold added to the average of the six pixels above it. Candidates are retained only if they form groups larger than 130~pixels, suppressing isolated pixels or small groups. Given the sample's geometry, the edge pixels and all pixels below them are selected for the mask. Two binary dilation iterations are then applied. Finally, Gaussian smoothing (radius~2, $\sigma=2$) is applied to the mask, converting the binary, sharp-edged mask into one with smooth boundaries.

\textbf{Row equalization.} At any given time step, the sum of each row of pixels is assumed to be proportional to the projected material mass of the corresponding slice of the sample. Because the mass of that slice is invariant to projection direction, the row-sum phase shift should approximately match across detectors at the same row height. With the additional assumption that the melt pool’s effect on row phase shifts is small compared with imaging errors, this holds for all time steps. We therefore compute, at each height, the average row sum over all time steps and detectors, and scale each detector’s row to this average. The scaling factor is bounded to at most~5 to avoid unreasonably large corrections, which could otherwise occur in rows with few nonzero pixels.

\textbf{Particle masking.} The detector~3 projections contain a powder particle traversing the melt-pool region. To prevent reconstruction artifacts, a binary mask sequence is generated to exclude the affected pixels from the training rays.

\subsection{Adapted X-Hexplane for rotation-XMPI}

The reconstruction method used in this work is based on STRT and, specifically, the 4D reconstruction framework X-Hexplane~\cite{hu2025super}. X-Hexplane leverages the Hexplane representation~\cite{cao2023hexplane} of 4D dynamical data via tensor factorization and implements X-ray physics based on the projection approximation to model the X-ray interaction with the sample. Self-supervised training infers the reconstruction purely from the measured projections without any pretraining.

The difference between rotation-XMPI acquisition and limited-angle tomography is illustrated in Fig.~\ref{fig:STRTvsXMPI}. In the rotation-XMPI case, a reconstruction is not assigned projections from different times at which the sample’s position or geometry may have changed. Moreover, although only one imaging time step contributes to one rotation-XMPI reconstruction time step, each reconstruction is formed from a set of projections spanning a wider angular range than in single-detector limited-angle tomography. The following main changes are therefore implemented in the reconstruction algorithm:
\begin{itemize}
\item Instead of one detector, the XMPI setup projects X-rays onto three detector scintillators from different directions simultaneously.
\item Because three distinct view directions are available, the reconstruction does not merge multiple imaging time steps into one reconstruction time step. One projection time step yields one reconstruction time step, increasing the reconstruction frame rate.
\end{itemize}

To compare tomography, STRT, and rotation-XMPI, the pure STRT algorithm is evaluated with 1, 10, 50, and 100 projection time steps per reconstruction time step while using only a single detector’s data rather than all detectors. The settings for rotation-XMPI and STRT reconstructions are listed in the supplementary information and were held constant for both reconstructions to allow direct comparability.

\subsection{Normalized cross-correlation (NCC)}
Let $A, B \in \mathbf{R}^{N_x \times N_y \times N_z}$ denote two 3D image volumes, and let $N = N_x N_y N_z$ be the total number of voxels. The volumes were flattened into vectors $a, b \in \mathbf{R}^{N}$ for evaluation.

The normalized cross-correlation (NCC) between $A$ and $B$ was computed as
\begin{equation}
\mathrm{NCC} =
\frac{\displaystyle \sum_{i=1}^{N} (a_i - \bar{a})(b_i - \bar{b})}
{\displaystyle \sqrt{\sum_{i=1}^{N} (a_i - \bar{a})^2}
\sqrt{\sum_{i=1}^{N} (b_i - \bar{b})^2}},
\end{equation}
where $\bar{a}$ and $\bar{b}$ denote the mean voxel intensities of $a$ and $b$, respectively.

\section{Data Availability}
The datasets generated and analyzed during the current study will be made publicly available upon publication of the peer-reviewed article.

\section{Code Availability}
The X-Hexplane reconstruction and the preprocessing pipeline code will be made publicly available upon publication of the peer-reviewed article.

\section{Acknowledgments}
This work was funded by SNF under Grant Number 228275 (Project 4DIAM) and ERC-2020-STG 3DX-FLASH 948426. We acknowledge the MAX IV Laboratory for beamtime on the ForMAX beamline under proposal 20241696. Research conducted at MAX IV, a Swedish national user facility, is supported by Vetenskapsrådet (Swedish Research Council, VR) under contract 2018-07152, Vinnova (Swedish Governmental Agency for Innovation Systems) under contract 2018-04969, and Formas under contract 2019-02496. Revisions to this work were made using large language models solely on text formulation to enhance readability and grammar. Sections of the code were written with the help of an LLM, then reviewed, edited, and tested.

\section{Author contributions}
L.W.: Investigation, Methodology, Software, Data curation, Formal analysis, Visualization, Writing – original draft.
E.J.: Investigation, Methodology, Data curation, Writing – review \& editing.
Z.H.: Software, Methodology, Data curation, Investigation.
Z.Y.: Software, Methodology, Data curation, Investigation.
E.M.A.: Data curation, Methodology, Investigation.
J.K.R.: Data curation, Methodology, Investigation.
Y.Z.: Software, Methodology, Investigation.
M.G.M.: Conceptualization, Methodology, Data curation, Investigation, Supervision, Writing - review \& editing, Project administration, Funding acquisition.
M.B.: Supervision, Writing - review \& editing, Project administration, Funding acquisition.
M.A.: Methodology, Supervision, Visualization, Writing – review \& editing, Project administration, Funding acquisition.
P.V.P.: Villanueva Perez: Conceptualization, Methodology, Data curation, Software, Investigation, Supervision, Writing – review \& editing, Project administration, Funding acquisition.

\section{Competing interests}
The authors declare no competing interests.


\begin{thebibliography}{10}
\expandafter\ifx\csname url\endcsname\relax
  \def\url#1{\texttt{#1}}\fi
\expandafter\ifx\csname urlprefix\endcsname\relax\def\urlprefix{URL }\fi
\providecommand{\bibinfo}[2]{#2}
\providecommand{\eprint}[2][]{\url{#2}}

\bibitem{khairallah2016laser}
\bibinfo{author}{Khairallah, S.~A.}, \bibinfo{author}{Anderson, A.~T.}, \bibinfo{author}{Rubenchik, A.} \& \bibinfo{author}{King, W.~E.}
\newblock \bibinfo{title}{Laser powder-bed fusion additive manufacturing: Physics of complex melt flow and formation mechanisms of pores, spatter, and denudation zones}.
\newblock \emph{\bibinfo{journal}{Acta Materialia}} \textbf{\bibinfo{volume}{108}}, \bibinfo{pages}{36--45} (\bibinfo{year}{2016}).

\bibitem{khairallah2020controlling}
\bibinfo{author}{Khairallah, S.~A.} \emph{et~al.}
\newblock \bibinfo{title}{Controlling interdependent meso-nanosecond dynamics and defect generation in metal 3d printing}.
\newblock \emph{\bibinfo{journal}{Science}} \textbf{\bibinfo{volume}{368}}, \bibinfo{pages}{660--665} (\bibinfo{year}{2020}).

\bibitem{gan2021universal}
\bibinfo{author}{Gan, Z.} \emph{et~al.}
\newblock \bibinfo{title}{Universal scaling laws of keyhole stability and porosity in 3d printing of metals}.
\newblock \emph{\bibinfo{journal}{Nature communications}} \textbf{\bibinfo{volume}{12}}, \bibinfo{pages}{2379} (\bibinfo{year}{2021}).

\bibitem{afrasiabi2023modelling}
\bibinfo{author}{Afrasiabi, M.} \& \bibinfo{author}{Bambach, M.}
\newblock \bibinfo{title}{Modelling and simulation of metal additive manufacturing processes with particle methods: A review}.
\newblock \emph{\bibinfo{journal}{Virtual and Physical Prototyping}} \textbf{\bibinfo{volume}{18}}, \bibinfo{pages}{e2274494} (\bibinfo{year}{2023}).

\bibitem{ullah2025additive}
\bibinfo{author}{Ullah, A.} \emph{et~al.}
\newblock \bibinfo{title}{Additive manufacturing of ceramics via the laser powder bed fusion process}.
\newblock \emph{\bibinfo{journal}{International journal of applied ceramic technology}} \textbf{\bibinfo{volume}{22}}, \bibinfo{pages}{e15087} (\bibinfo{year}{2025}).

\bibitem{muther2025identifying}
\bibinfo{author}{Muther, A.} \emph{et~al.}
\newblock \bibinfo{title}{Identifying melt pool behavior in ceramics pbf-lb via operando synchrotron tomographic microscopy and high-fidelity process modeling}.
\newblock \emph{\bibinfo{journal}{Additive Manufacturing}} \textbf{\bibinfo{volume}{103}}, \bibinfo{pages}{104756} (\bibinfo{year}{2025}).

\bibitem{luthi2023adaptive}
\bibinfo{author}{L{\"u}thi, C.}, \bibinfo{author}{Afrasiabi, M.} \& \bibinfo{author}{Bambach, M.}
\newblock \bibinfo{title}{An adaptive smoothed particle hydrodynamics (sph) scheme for efficient melt pool simulations in additive manufacturing}.
\newblock \emph{\bibinfo{journal}{Computers \& Mathematics with Applications}} \textbf{\bibinfo{volume}{139}}, \bibinfo{pages}{7--27} (\bibinfo{year}{2023}).

\bibitem{moges2019review}
\bibinfo{author}{Moges, T.}, \bibinfo{author}{Ameta, G.} \& \bibinfo{author}{Witherell, P.}
\newblock \bibinfo{title}{A review of model inaccuracy and parameter uncertainty in laser powder bed fusion models and simulations}.
\newblock \emph{\bibinfo{journal}{Journal of manufacturing science and engineering}} \textbf{\bibinfo{volume}{141}}, \bibinfo{pages}{040801} (\bibinfo{year}{2019}).

\bibitem{parab2018ultrafast}
\bibinfo{author}{Parab, N.~D.} \emph{et~al.}
\newblock \bibinfo{title}{Ultrafast x-ray imaging of laser--metal additive manufacturing processes}.
\newblock \emph{\bibinfo{journal}{Synchrotron Radiation}} \textbf{\bibinfo{volume}{25}}, \bibinfo{pages}{1467--1477} (\bibinfo{year}{2018}).

\bibitem{martin2019dynamics}
\bibinfo{author}{Martin, A.~A.} \emph{et~al.}
\newblock \bibinfo{title}{Dynamics of pore formation during laser powder bed fusion additive manufacturing}.
\newblock \emph{\bibinfo{journal}{Nature communications}} \textbf{\bibinfo{volume}{10}}, \bibinfo{pages}{1987} (\bibinfo{year}{2019}).

\bibitem{martin2022laser}
\bibinfo{author}{Martin, A.~A.} \emph{et~al.}
\newblock \bibinfo{title}{A laser powder bed fusion system for operando synchrotron x-ray imaging and correlative diagnostic experiments at the stanford synchrotron radiation lightsource}.
\newblock \emph{\bibinfo{journal}{Review of Scientific Instruments}} \textbf{\bibinfo{volume}{93}} (\bibinfo{year}{2022}).

\bibitem{garcia2021tomoscopy}
\bibinfo{author}{Garc{\'\i}a-Moreno, F.} \emph{et~al.}
\newblock \bibinfo{title}{Tomoscopy: Time-resolved tomography for dynamic processes in materials}.
\newblock \emph{\bibinfo{journal}{Advanced Materials}} \textbf{\bibinfo{volume}{33}}, \bibinfo{pages}{2104659} (\bibinfo{year}{2021}).

\bibitem{garcia2023x}
\bibinfo{author}{Garc{\'\i}a-Moreno, F.}, \bibinfo{author}{Neu, T.~R.}, \bibinfo{author}{Kamm, P.~H.} \& \bibinfo{author}{Banhart, J.}
\newblock \bibinfo{title}{X-ray tomography and tomoscopy on metals: A review}.
\newblock \emph{\bibinfo{journal}{Advanced Engineering Materials}} \textbf{\bibinfo{volume}{25}}, \bibinfo{pages}{2201355} (\bibinfo{year}{2023}).

\bibitem{makowska2023operando}
\bibinfo{author}{Makowska, M.~G.} \emph{et~al.}
\newblock \bibinfo{title}{Operando tomographic microscopy during laser-based powder bed fusion of alumina}.
\newblock \emph{\bibinfo{journal}{Communications Materials}} \textbf{\bibinfo{volume}{4}}, \bibinfo{pages}{73} (\bibinfo{year}{2023}).

\bibitem{hoshino2011stereo}
\bibinfo{author}{Hoshino, M.} \emph{et~al.}
\newblock \bibinfo{title}{Development of an x-ray real-time stereo imaging technique using synchrotron radiation}.
\newblock \emph{\bibinfo{journal}{Journal of Synchrotron Radiation}} \textbf{\bibinfo{volume}{18}}, \bibinfo{pages}{569--574} (\bibinfo{year}{2011}).
\newblock \bibinfo{note}{Publisher: International Union of Crystallography}.

\bibitem{mokso2015StereoDual}
\bibinfo{author}{Mokso, R.} \& \bibinfo{author}{Oberta, P.}
\newblock \bibinfo{title}{Simultaneous dual-energy x-ray stereo imaging}.
\newblock \emph{\bibinfo{journal}{Journal of Synchrotron Radiation}} \textbf{\bibinfo{volume}{22}}, \bibinfo{pages}{1078--1082} (\bibinfo{year}{2015}).

\bibitem{Villanueva-Perez2018XMPI}
\bibinfo{author}{Villanueva-Perez, P.} \emph{et~al.}
\newblock \bibinfo{title}{Hard x-ray multi-projection imaging for single-shot approaches}.
\newblock \emph{\bibinfo{journal}{Optica}} \textbf{\bibinfo{volume}{5}}, \bibinfo{pages}{1521--1524} (\bibinfo{year}{2018}).
\newblock \bibinfo{note}{Publisher: Optical Society of America}.

\bibitem{Duarte2019CDIstereo}
\bibinfo{author}{Duarte, J.} \emph{et~al.}
\newblock \bibinfo{title}{Computed stereo lensless x-ray imaging}.
\newblock \emph{\bibinfo{journal}{Nature Photonics}} \textbf{\bibinfo{volume}{13}}, \bibinfo{pages}{449--453} (\bibinfo{year}{2019}).

\bibitem{voegeli2020multibeamWhite}
\bibinfo{author}{Voegeli, W.} \emph{et~al.}
\newblock \bibinfo{title}{Multibeam x-ray optical system for high-speed tomography}.
\newblock \emph{\bibinfo{journal}{Optica}} \textbf{\bibinfo{volume}{7}}, \bibinfo{pages}{514--517} (\bibinfo{year}{2020}).

\bibitem{asimakopoulou2024kHzXMPI}
\bibinfo{author}{Asimakopoulou, E.~M.} \emph{et~al.}
\newblock \bibinfo{title}{Development towards high-resolution {kHz}-speed rotation-free volumetric imaging}.
\newblock \emph{\bibinfo{journal}{Optics Express}} \textbf{\bibinfo{volume}{32}}, \bibinfo{pages}{4413--4426} (\bibinfo{year}{2024}).
\newblock \bibinfo{note}{Publisher: Optica Publishing Group}.

\bibitem{voegeli2024multibeamSiArray}
\bibinfo{author}{Voegeli, W.} \emph{et~al.}
\newblock \bibinfo{title}{Multibeam x-ray tomography optical system for narrow-energy-bandwidth synchrotron radiation}.
\newblock \emph{\bibinfo{journal}{Applied Physics Express}} \textbf{\bibinfo{volume}{17}}, \bibinfo{pages}{032002} (\bibinfo{year}{2024}).

\bibitem{villanueva2023megahertz}
\bibinfo{author}{Villanueva-Perez, P.} \emph{et~al.}
\newblock \bibinfo{title}{Megahertz x-ray multi-projection imaging}.
\newblock \emph{\bibinfo{journal}{arXiv preprint arXiv:2305.11920}}  (\bibinfo{year}{2023}).

\bibitem{Bellucci2024CrystalsXMPI}
\bibinfo{author}{Bellucci, V.} \emph{et~al.}
\newblock \bibinfo{title}{{Development of crystal optics for X-ray multi-projection imaging for synchrotron and XFEL sources}}.
\newblock \emph{\bibinfo{journal}{Journal of Synchrotron Radiation}} \textbf{\bibinfo{volume}{31}}, \bibinfo{pages}{1534--1550} (\bibinfo{year}{2024}).

\bibitem{goethals2022dynamic}
\bibinfo{author}{Goethals, W.} \emph{et~al.}
\newblock \bibinfo{title}{Dynamic ct reconstruction with improved temporal resolution for scanning of fluid flow in porous media}.
\newblock \emph{\bibinfo{journal}{Water Resources Research}} \textbf{\bibinfo{volume}{58}}, \bibinfo{pages}{e2021WR031365} (\bibinfo{year}{2022}).

\bibitem{mohan2015timbir}
\bibinfo{author}{Mohan, K.~A.} \emph{et~al.}
\newblock \bibinfo{title}{Timbir: A method for time-space reconstruction from interlaced views}.
\newblock \emph{\bibinfo{journal}{IEEE Transactions on Computational Imaging}} \textbf{\bibinfo{volume}{1}}, \bibinfo{pages}{96--111} (\bibinfo{year}{2015}).

\bibitem{kazantsev20154d}
\bibinfo{author}{Kazantsev, D.} \emph{et~al.}
\newblock \bibinfo{title}{4d-ct reconstruction with unified spatial-temporal patch-based regularization}.
\newblock \emph{\bibinfo{journal}{Inverse problems and imaging}} \textbf{\bibinfo{volume}{9}}, \bibinfo{pages}{447--467} (\bibinfo{year}{2015}).

\bibitem{goethals2025dyrect}
\bibinfo{author}{Goethals, W.}, \bibinfo{author}{Bultreys, T.}, \bibinfo{author}{Berg, S.}, \bibinfo{author}{Boone, M.~N.} \& \bibinfo{author}{Aelterman, J.}
\newblock \bibinfo{title}{Dyrect computed tomography: Dynamic reconstruction of events on a continuous timescale}.
\newblock \emph{\bibinfo{journal}{IEEE Transactions on Computational Imaging}} \textbf{\bibinfo{volume}{11}}, \bibinfo{pages}{638--649} (\bibinfo{year}{2025}).

\bibitem{mildenhall2021nerf}
\bibinfo{author}{Mildenhall, B.} \emph{et~al.}
\newblock \bibinfo{title}{Nerf: Representing scenes as neural radiance fields for view synthesis}.
\newblock \emph{\bibinfo{journal}{Communications of the ACM}} \textbf{\bibinfo{volume}{65}}, \bibinfo{pages}{99--106} (\bibinfo{year}{2021}).

\bibitem{zhang20254d}
\bibinfo{author}{Zhang, Y.}, \bibinfo{author}{Yao, Z.}, \bibinfo{author}{Kl{\"o}fkorn, R.}, \bibinfo{author}{Ritschel, T.} \& \bibinfo{author}{Villanueva-Perez, P.}
\newblock \bibinfo{title}{4d-onix for reconstructing 3d movies from sparse x-ray projections via deep learning}.
\newblock \emph{\bibinfo{journal}{Communications Engineering}} \textbf{\bibinfo{volume}{4}}, \bibinfo{pages}{1--12} (\bibinfo{year}{2025}).

\bibitem{reed2021dynamic}
\bibinfo{author}{Reed, A.~W.} \emph{et~al.}
\newblock \bibinfo{title}{Dynamic ct reconstruction from limited views with implicit neural representations and parametric motion fields}.
\newblock In \emph{\bibinfo{booktitle}{Proceedings of the IEEE/CVF International Conference on Computer Vision}}, \bibinfo{pages}{2258--2268} (\bibinfo{year}{2021}).

\bibitem{hu2025super}
\bibinfo{author}{Hu, Z.} \emph{et~al.}
\newblock \bibinfo{title}{Super time-resolved tomography}.
\newblock \emph{\bibinfo{journal}{Advanced Science}} \textbf{\bibinfo{volume}{13}}, \bibinfo{pages}{e11933} (\bibinfo{year}{2026}).

\bibitem{cao2023hexplane}
\bibinfo{author}{Cao, A.} \& \bibinfo{author}{Johnson, J.}
\newblock \bibinfo{title}{Hexplane: A fast representation for dynamic scenes}.
\newblock In \emph{\bibinfo{booktitle}{2023 IEEE/CVF Conference on Computer Vision and Pattern Recognition (CVPR)}}, \bibinfo{pages}{130--141} (\bibinfo{year}{2023}).

\bibitem{luthi2025multi}
\bibinfo{author}{L{\"u}thi, C.}, \bibinfo{author}{Bambach, M.} \& \bibinfo{author}{Afrasiabi, M.}
\newblock \bibinfo{title}{Multi-3: A gpu-enhanced meshfree simulation framework for multi-track, multi-layer, and multi-material laser powder bed fusion processes}.
\newblock \emph{\bibinfo{journal}{Journal of Manufacturing Processes}} \textbf{\bibinfo{volume}{147}}, \bibinfo{pages}{29--48} (\bibinfo{year}{2025}).

\bibitem{afrasiabi2022effect}
\bibinfo{author}{Afrasiabi, M.}, \bibinfo{author}{Keller, D.}, \bibinfo{author}{L{\"u}thi, C.}, \bibinfo{author}{Bambach, M.} \& \bibinfo{author}{Wegener, K.}
\newblock \bibinfo{title}{Effect of process parameters on melt pool geometry in laser powder bed fusion of metals: a numerical investigation}.
\newblock \emph{\bibinfo{journal}{Procedia CIRP}} \textbf{\bibinfo{volume}{113}}, \bibinfo{pages}{378--384} (\bibinfo{year}{2022}).

\bibitem{rogalinski2025time}
\bibinfo{author}{Rogalinski, J.~K.} \emph{et~al.}
\newblock \bibinfo{title}{{Time-resolved 3D imaging opportunities with XMPI at ForMAX}}.
\newblock \emph{\bibinfo{journal}{Journal of Synchrotron Radiation}} \textbf{\bibinfo{volume}{33}} (\bibinfo{year}{2026}).
\newblock \urlprefix\url{https://doi.org/10.1107/S1600577525011038}.

\bibitem{authier2006dynamical}
\bibinfo{author}{Authier, A.}
\newblock \bibinfo{title}{Dynamical theory of x-ray diffraction}.
\newblock In \emph{\bibinfo{booktitle}{International tables for crystallography volume B: reciprocal space}}, \bibinfo{pages}{534--551} (\bibinfo{publisher}{Springer}, \bibinfo{year}{2006}).

\bibitem{dowd1999developments}
\bibinfo{author}{Dowd, B.~A.} \emph{et~al.}
\newblock \bibinfo{title}{Developments in synchrotron x-ray computed microtomography at the national synchrotron light source}.
\newblock In \emph{\bibinfo{booktitle}{Developments in X-ray Tomography II}}, vol. \bibinfo{volume}{3772}, \bibinfo{pages}{224--236} (\bibinfo{organization}{SPIE}, \bibinfo{year}{1999}).

\bibitem{gursoy2014tomopy}
\bibinfo{author}{G{\"u}rsoy, D.}, \bibinfo{author}{De~Carlo, F.}, \bibinfo{author}{Xiao, X.} \& \bibinfo{author}{Jacobsen, C.}
\newblock \bibinfo{title}{Tomopy: a framework for the analysis of synchrotron tomographic data}.
\newblock \emph{\bibinfo{journal}{Synchrotron Radiation}} \textbf{\bibinfo{volume}{21}}, \bibinfo{pages}{1188--1193} (\bibinfo{year}{2014}).

\bibitem{yao2025physics-informed4D}
\bibinfo{author}{Yao, Z.} \emph{et~al.}
\newblock \bibinfo{title}{Physics-informed {4D} x-ray image reconstruction from ultra-sparse spatiotemporal data}.
\newblock \emph{\bibinfo{journal}{Measurement Science and Technology}} \textbf{\bibinfo{volume}{36}}, \bibinfo{pages}{085403} (\bibinfo{year}{2025}).
\newblock \bibinfo{note}{Publisher: IOP Publishing}.

\bibitem{nygaard2024formax}
\bibinfo{author}{Nyg{\aa}rd, K.} \emph{et~al.}
\newblock \bibinfo{title}{{ForMAX {--} a beamline for multiscale and multimodal structural characterization of hierarchical materials}}.
\newblock \emph{\bibinfo{journal}{Journal of Synchrotron Radiation}} \textbf{\bibinfo{volume}{31}}, \bibinfo{pages}{363--377} (\bibinfo{year}{2024}).

\bibitem{donoho1994ideal}
\bibinfo{author}{Donoho, D.~L.} \& \bibinfo{author}{Johnstone, I.~M.}
\newblock \bibinfo{title}{Ideal spatial adaptation by wavelet shrinkage}.
\newblock \emph{\bibinfo{journal}{biometrika}} \textbf{\bibinfo{volume}{81}}, \bibinfo{pages}{425--455} (\bibinfo{year}{1994}).

\bibitem{darbon2008fast}
\bibinfo{author}{Darbon, J.}, \bibinfo{author}{Cunha, A.}, \bibinfo{author}{Chan, T.~F.}, \bibinfo{author}{Osher, S.} \& \bibinfo{author}{Jensen, G.~J.}
\newblock \bibinfo{title}{Fast nonlocal filtering applied to electron cryomicroscopy}.
\newblock In \emph{\bibinfo{booktitle}{2008 5th IEEE International Symposium on biomedical imaging: from nano to macro}}, \bibinfo{pages}{1331--1334} (\bibinfo{organization}{IEEE}, \bibinfo{year}{2008}).

\bibitem{paganin2002simultaneous}
\bibinfo{author}{Paganin, D.}, \bibinfo{author}{Mayo, S.~C.}, \bibinfo{author}{Gureyev, T.~E.}, \bibinfo{author}{Miller, P.~R.} \& \bibinfo{author}{Wilkins, S.~W.}
\newblock \bibinfo{title}{Simultaneous phase and amplitude extraction from a single defocused image of a homogeneous object}.
\newblock \emph{\bibinfo{journal}{Journal of microscopy}} \textbf{\bibinfo{volume}{206}}, \bibinfo{pages}{33--40} (\bibinfo{year}{2002}).

\end{thebibliography}
\end{document}